\documentclass[a4paper]{mem}
\usepackage{natbib}
\usepackage{graphicx}
\usepackage[a4paper]{hyperref}
\idline{73}{1}
\def\pdot {\dot P}

\def\ltsima{$\; \buildrel < \over \sim \;$}
\def\lsim{\lower.5ex\hbox{\ltsima}}

\def\gsim{\lower.5ex\hbox{\gtsima}}

\def\ee {1E~1207.4--5209~}
\def\cha {\textit{Chandra~}}

\def\xmm  {\textit{XMM-Newton}}

\begin{document}
   \title{1E1207.4-5209 - a Unique Object
}

   \author{G.F. Bignami \inst{1,2}, A. De Luca \inst{3,4}, P.A. Caraveo\inst{3}, S. Mereghetti\inst{3}, 
   M. Moroni\inst{3} \and R.P.Mignani\inst{5}}
          

   \offprints{G.F. Bignami}

   \institute{Centre d'Etude Spatiale des Rayonnements, CNRS-UPS, 9, Avenue du Colonel Roche, 31028, Toulouse Cedex 4, France \email{bignami@cesr.fr}\\ 
              \and  Universit\`a degli Studi di Pavia, Dipartimento di Fisica Nucleare e Teorica, Via Bassi 6, 27100 Pavia, Italy\\
              \and INAF/IASF ``G. Occhialini'', Via Bassini 15, 20133 Milano, Italy\\
              \and Universit\`a di Milano Bicocca, Dipartimento di Fisica, P.za della Scienza 3, 20126 Milano, Italy\\
              \and ESO, Karl Schwarzschild Strasse 2, D-85740, Garching, Germany
                          }

   \abstract{
The discovery of deep spectral features in the X-ray spectrum of
1E1207.4-5209 focussed the attention of the astronomical community
on this radio-quiet NS, making it the most intensively observed INS ever.
The harvest of X-ray photons, collected mainly by XMM-Newton, unveiled
a number of unique characteristics, raising questions on this source very nature.

   }
   \authorrunning{G.F. Bignami et al. et al.}
   \titlerunning{Out of the chorus line}
   \maketitle
%

\section{Introduction}
Neutron star atmosphere models predicted the presence of
absorption features depending on  atmospheric composition, but high
quality spectra, collected both by \cha and by \xmm, did
not yield  evidence for any such feature (see Pavlov et al. 2002a
and Becker and Aschenbach 2002 for recent reviews). 
INS spectra are well fitted
by one or more black-body curves with, possibly, a power law
contribution at higher energies, but with no absorption or
emission features. \\ The spectrum of 1E1207.4-5209, on the
contrary,  appeared from the start as dominated by two broad absorption features seen, at
0.7 and 1.4 keV, both by \cha (Sanwal et al, 2002)  and
\xmm~(Mereghetti et al, 2002). To better understand the
nature of such features, \xmm~devoted two orbits 
to 1E1207.4-5209, for a total observing time of 257,303 sec. In the two
MOS EPIC cameras the source yielded  74,600 and 76,700 photons in
the energy range 0.2 - 3.5 keV, while the pn camera recorded
208,000 photons, time-tagged to allow for timing studies. Analysis
of this long observation, while confirming  the  two
phase-dependent absorption lines at 0.7 and 1.4 keV, unveiled a
statistically significant third line at $\sim$2.1 keV, as  well
as  a  possible fourth  feature at  2.8 keV.  The nearly perfect 1:2:3:4
ratio  of the line centroids,  as well as  the phase variation,
naturally following the pulsar B-field rotation, strongly suggest
that such lines are due to cyclotron resonance scattering
(Bignami et al. 2003).  A year after the \xmm~observation, \ee
was deeply scrutinized by Chandra for $\sim$300 ksec, thus logging an 
overall effective exposure slightly larger than that devoted 
to the most popular NSs so far, such as Crab, Vela and RX J1856.5-3754.

The release of a new, and improved, instrument calibration software
prompted us to revisit the \xmm~ data set.
Here we shall briefly report on some highlights of our new spectral and
temporal analysis (see De Luca et al. 2004 for details).
In addition, the ``best'' X-ray positional information is used, in 
conjunction with a new optical observation, to investigate the source 
optical behaviour.

\section{Timing Analysis}

\begin{figure*}
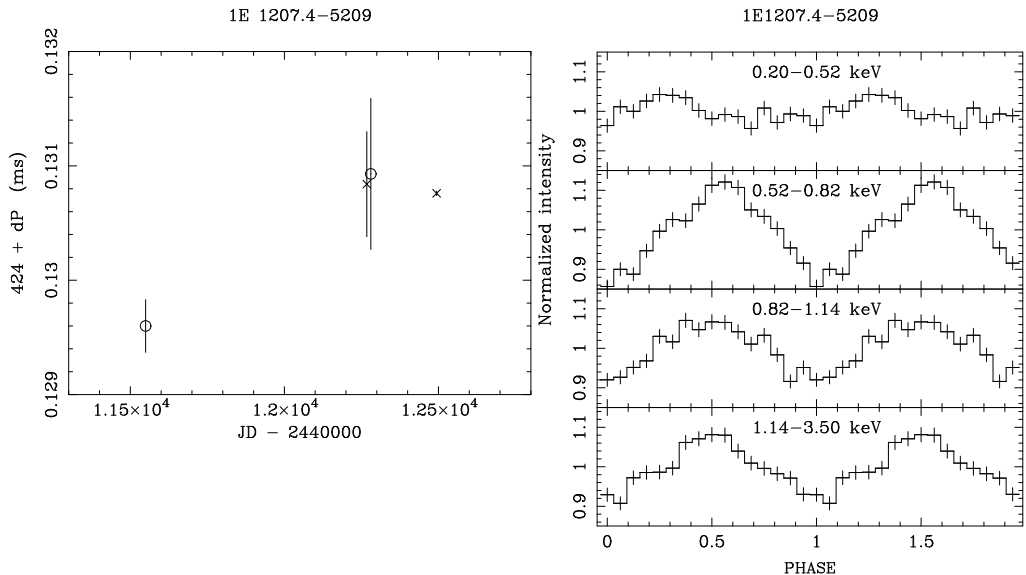

   \centering
   \resizebox{\hsize}{!}{\includegraphics[angle=-90,width=8.3cm]{bignami_fig1.ps}
   \includegraphics[angle=-90,width=7.7cm]{bignami_fig2.ps}}
     \caption{(Left) Period history of \ee . Circles are the \cha
measurements, crosses \xmm. (Right) Folded light curve of \ee in four energy ranges. Fitting
the pulse profiles with a function consisting of a constant plus a
sinusoid, we obtain pulsed fraction values of (2.7$\pm$0.7)\% in
the 0.2-0.52 keV energy range, (11.1$\pm$0.7)\% for 0.52-0.82 keV
, (7.1$\pm$0.7)\% for 0.82-1.14 keV, and (7.1$\pm$0.7)\% 1.14-3.5
keV.}
        \label{timing}
    \end{figure*}

After converting the arrival times of the 208,000 pn photons to
the Solar System Barycenter, we searched the period range from
424.12 to 424.14 ms using both a folding algorithm with 8 phase
bins and the Rayleigh  test. The best period value and its
uncertainty (P = 424.13076$\pm$0.00002 ms) were determined
following the procedure outlined in Mereghetti et al. (2002).
Comparing the new period measurement of \ee with that obtained
with \cha in January 2000 (Pavlov et al. 2002b), we obtain a
period derivative $\pdot$=(1.4$\pm$0.3)$\times$10$^{-14}$ s
s$^{-1}$. However, Fig.~\ref{timing} (left panel) shows that the $\pdot$ value rests
totally on the first  \cha period measurement. Using only the
3 most recent values, the period derivative is unconstrained.
Thus, we cannot exclude that the observed spin-down, based on
only a few sparse measurements, be affected by glitches or
Doppler shifts induced by orbital motion. Questioning the
object's $\pdot$ would have far reaching consequences for the
understanding of 1E1207.4-5209 since the serious discrepancy
between the pulsar characteritic age ($\tau_{c}\sim5\times10^5$ yrs)
and the SNR age ($\tau_{SNR}\sim7$ kyrs) is entirely
based on the value inferred from the measurements summarized in
Fig.~\ref{timing} (left panel).

To study the energy dependence of the pulse profile, we divided
the data in four channels with approximately 52,000 counts each:
0.2-0.52 keV, 0.52-0.82 keV, 0.82-1.14 keV and 1.14-3.5 keV. The
pulse profiles in the different energy ranges (Fig.~\ref{timing}, right panel)   show a
broad, nearly sinusoidal shape, with a pulsed fraction varying
from $\sim$ 3 to $\sim$ 11 \%  in the four energy intervals.  It
is worth noting that the minimum pulsed fraction is found in the
0.20-0.52 keV energy range, the only portion of the spectrum free
from absorption lines. Indeed, Fig.~\ref{timing} (right panel) is an independent
confirmation of the findings of Bignami et al (2003) who ascribed
the source pulsation to the absorption lines phase variation.

Finally, comparing the shapes of the light curves of Fig.~\ref{timing} (right panel), we
see for the first time a phase shift of nearly 90$^{\circ}$
between the profile in the lowest energy range ($<$0.52 keV) and
those at higher energies.

\section{Spectral analysis}

   \begin{figure*}
   \centering
  \includegraphics[angle=-90,width=11cm]{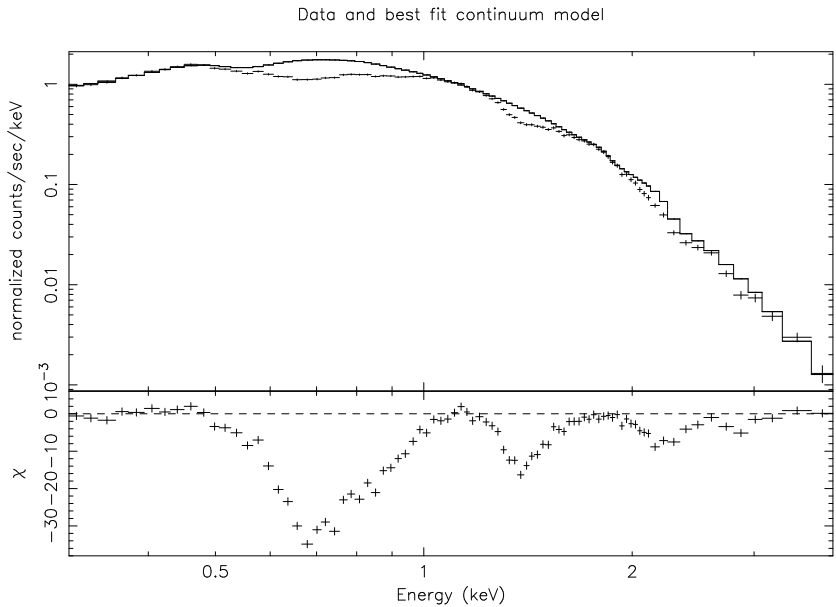}
   \caption{The spectrum of 1E 1207.4-5209 as seen by the pn camera. The upper panel shows
   the data points together with the best fitting continuum model folded with the
   instrumental response. The lower panel shows the residuals in units of standard deviations from
   the best fitting continuum. The presence of four absorption features at $\sim0.7$, $\sim1.4$, 
   $\sim2.1$, $\sim2.8$ keV is evident.}
              \label{spectrum}%
    \end{figure*}

   \begin{figure*}
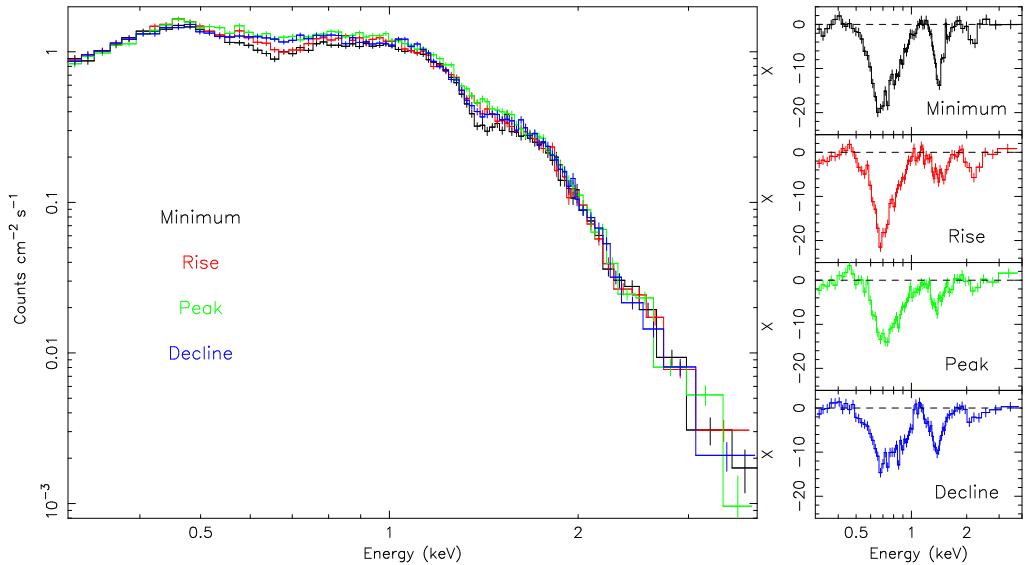

   \centering
   \resizebox{\hsize}{!}{\includegraphics[angle=-90,clip=true]{bignami_fig4.ps}
   \includegraphics[angle=-90,clip=true]{bignami_fig5.ps}}
     \caption{(Left) Comparison of the four phase-resolved spectra extracted from pn data.
   Colours indicate the phase-interval: black, minimum; red, rise; green, peak;
   blue, decline. The peak of the total light curve corresponds to the phase interval
   where the absorption features are at their minimum, while the light curve
   minimum happens when the absorption features are more important. (Right) Residuals
   in units of sigma obtained by comparing the data with the best fit continuum model 
   for the phase-resolved
   spectra. The pulse phase variations in width, depth and shape of the
   absorption features are evident.}
        \label{phaseres}
    \end{figure*}

The improved understanding of the instruments (concerning, e.g.,
its Quantum Efficiency, its Charge Transfer Inefficiency, and its redistribution
function) implemented in the most recent SAS release used here (SAS v5.4.1)
yields rather significant differences in the low energy (E$<$0.6 keV) part of the spectrum
with respect to our previous analysis (Bignami et al. 2003). This allowed for an update
of the best fitting parameters for both the continuum and the lines. 
The reader is referred to De Luca et al. (2004) for a complete discussion of the
spectral analysis of the three EPIC cameras. Here we shall focus on the pn data
set, which contributes more than half of the source statistics.
The best fitting continuum model comes from the sum of two blackbody functions. 
The cooler one has a temperature kT=$0.163\pm0.003$ keV and an emitting radius R=$4.6\pm0.1$ km, 
while the hotter one has kT=$0.319\pm0.002$ and R=$0.83\pm0.03$ km. 
Four absorption features are clearly seen in the
spectrum of \ee (see Fig~\ref{spectrum}) at the
harmonically spaced energies of
$\sim$0.7 keV, $\sim$1.4 keV, $\sim$2.1 keV and
$\sim$2.8 keV. The different spectral continuum model,
resulting from the improved calibrations, yields also  a more significant
detection of the third and fourth features with respect to our previous analysis.
 Using a simple gaussian in
absorption, we estimated with an F-test that the 2.1 keV
and the 2.8 keV features have a chance occurrence probability of
$\sim$10$^{-9}$ and  $\sim 10^{-3}$, respectively.

As a further step, we have studied the variations of the spectrum of \ee 
as a function of the pulse phase.
Following Bignami et al. (2003), we selected the phase
intervals corresponding to the peak (phase interval 0.40-0.65
with respect to Fig.~\ref{timing}, right panel), the declining
part (phase 0.65-0.90), the minimum (phase 0.90-1.15) and the
rising part (phase 0.15-0.40) of the folded light curve.
The resulting spectra (see Fig.~\ref{phaseres}) were fitted allowing both the thermal
continuum and the lines to vary.
The main results  can be summarized as follows:
\begin{itemize}
\item The two components of the continuum vary slightly 
both in temperature and in flux with the pulse phase, accounting at most for $\sim$3-5\%
of the source pulsation. An inspection of Fig.~\ref{phaseres}
clearly shows that the phase variation of
the features is largely responsible for the observed pulsation of the source.
\item the features are strongly dependent on the pulse phase and show significant 
variations in width, depth and shape 
The central energy of the 0.7 keV feature varies by 6\%
at most, while displacements of the other features are not
significant. The 2.8 keV feature is only marginally detected 
during the minimum and the rise intervals.
\item The relative intensity of the first three features varies with the phase.
\end{itemize}

\section{Optimizing the X-ray position}

To derive the sky coordinates of 1E 1207.4-5209, we computed independently
for the MOS1 and MOS2 cameras the boresight correction to be
applied to the default EPIC astrometry. We used the Guide Star
Catalog II
(GSC-II\footnote{http://www-gsss.stsci.edu/gsc/gsc2/GSC2home.htm})
to select, amongst our $\sim$200 serendipitous detections, 6
sources with a stellar counterpart to be used to correct the EPIC
astrometry. The rms error between the refined X-ray and GSC-II
positions is $\sim$1 arcsec per coordinate. The resulting MOS1
position of \ee is $\alpha_{J2000}=12h10m00.91s$,
$\delta_{J2000}=-52^{\circ}26'28.8"$ with an overall error radius
of 1.5 arcsec. The MOS2 position is $\alpha_{J2000}=12h10m00.84s$,
$\delta_{J2000}=-52^{\circ}26'27.6"$, also with an uncertainty of 1.5
arcsec, fully consistent with the MOS1 coordinates.

   \begin{figure}
   \centering
   \includegraphics[angle=0,width=6.5cm]{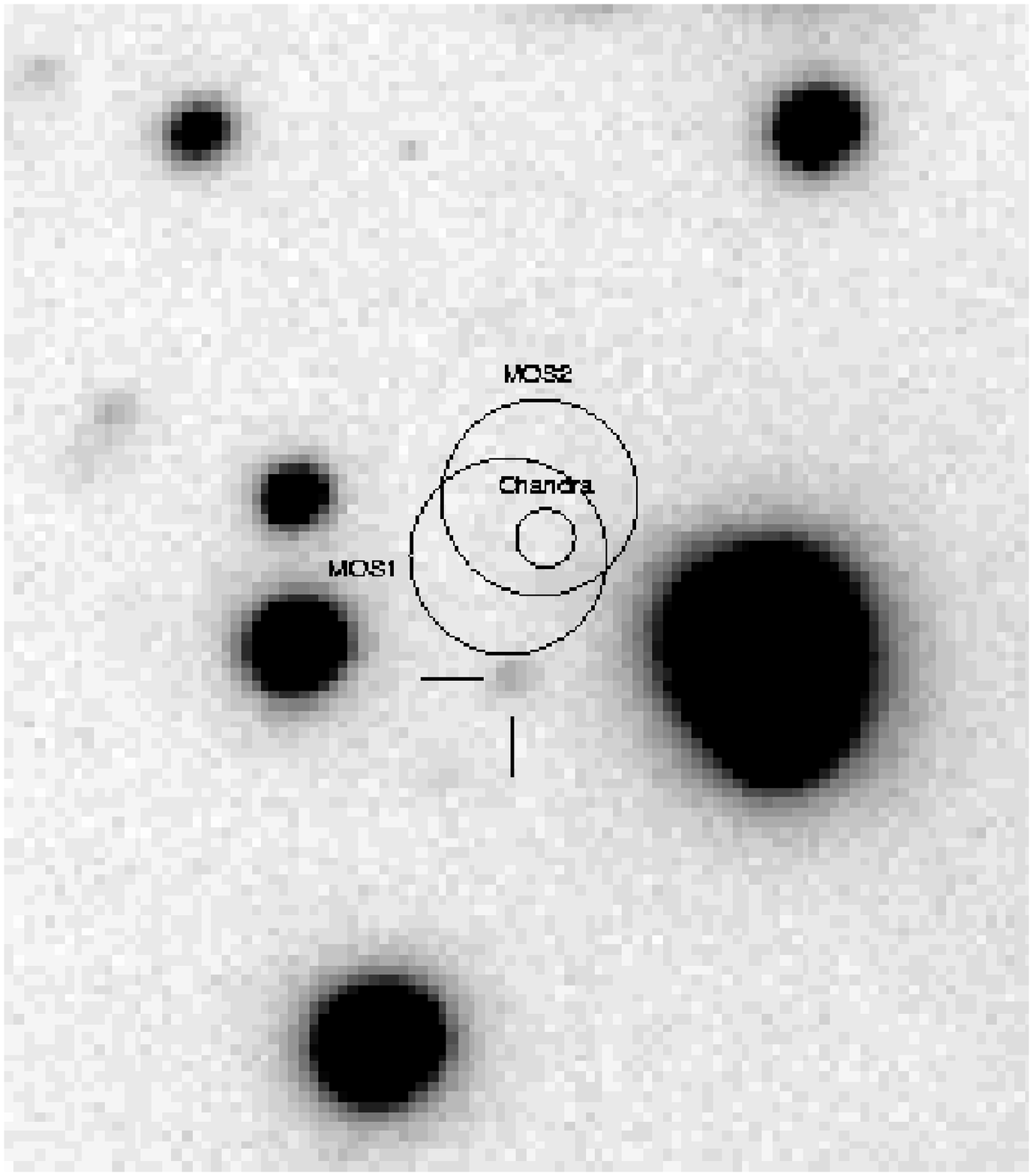}
      \caption{FORS1 $V$-band image centered on the  target position. The error circles from Chandra/ACIS (0.6$''$ radius),
    MOS1 and MOS2 (2 arcsec radius) have been superimposed.}
         \label{vlt}
   \end{figure}

In order to obtain an independent measurement on the position
of 1E 1207.4-5209, we have retrieved from the Chandra Data Archive a public
dataset relative to a recent (June 2003) ACIS observation 
(20 ksec in imaging mode) of the target.
As in the case of the EPIC data, we used the positions of
the serendipitous sources in the field to refine the astrometry
(see De Luca et al. 2004 for details).
The best Chandra/ACIS
position of \ee is $\alpha_{J2000}=12h10m00.826s$,
$\delta_{J2000}=-52^{\circ}26'28.43''$ with an uncertainty of
0.6$''$. The more accurate Chandra position lies inside the
intersection of the MOS1 and MOS2 error circles, confirming 
the correctness of the analysis and the absence
of systematics.

\section{Search for the optical counterpart}

The field of 1E 1207$-$5209 was observed with the 8.2-meter UT-1
Telescope (Antu) of the ESO VLT (Paranal Observatory).
Observations were  performed with the the FOcal Reducer and
Spectrograph 1 (FORS1) instrument. Images   were   acquired
through   the   Bessel   $V$ and $R$  filters  for  a  total
integration  time of  $\sim $2  and 3 hrs,  respectively.  Fig.~\ref{vlt}
 shows the inner portion  of the combined FORS1
$V$-band image centered on the target position, with the ACIS, MOS1 and
MOS2 error circles superimposed.    A faint object (marked with
the  two ticks in Fig.~\ref{vlt}) is detected just outside the southern
edge of the MOS1 error circle. It showed variability during the
time span covered by our observations. Also, its position falls  more
than 2 arcsec  away from the most probable region. For both reasons, we can rule it out as a
potential counterpart of \ee. \\No
candidate counterpart  is detected within the Chandra error circle 
(nor in the intersection of the MOS ones) 
down to R$\sim$27.1 and V$\sim$27.3, which we assume as
upper limits on the optical flux of 1E 1207$-$5209. 
For an  X-ray derived interstellar  absorption of $A_{V} =0.65$ at b$\sim10^{\circ}$ and a
distance of 2 kpc, our VLT u.l. rules out any hypothetical ``normal'' stellar 
companion (M$>$15 both in R and in V).  If we assume
that 1E  1207-5209 is indeed {\it  isolated}, we can  derive a neutron
star optical luminosity $\le 3.4 ~ 10^{28}$ erg s$^{-1}$ or $\le 4.6 ~
10^{-6}$ of  its rotational energy loss,  a value similar  to those of
middle-aged INSs.  Since the VLT flux upper limits are $\ge
100$ higher  than the extrapolation of  the \xmm~ blackbody  (De Luca et
al.  2004), they do not constrain our object's optical spectrum.

\section{Conclusions}

\ee is a radio-quiet NS at the center of a well known SNR. However,
the longer we observe such an object, the harder it becomes to make it fit 
into a known class of the NS family. The long observations devoted to \ee
both by \cha and by \xmm~ have unveiled a number of unique and
somewhat contradictory characteristics that, at the moment, defy standard 
theoretical interpretations.

The first and outmost problem seems to be the interpretation of the period evolution.
The usual, simplistic, monotonic fit yields a $\dot{P}$=(1.4$\pm$0.3)$\times$10$^{-14}$ s s$^{-1}$
that, combined with the period value, makes \ee much too old ($>$ 50 times) 
for its SNR.
Moreover, the value of the star's magnetic field inferred  on the basis of the classical
dipole braking formula (B=2.6$\pm$0.3$\times10^{12}$ G) is significantly
different from the value obtained from the cyclotron absorption lines interpreted
both in terms of electrons (B$\sim8\times10^{10}$ G) as well as protons (B$\sim1.6\times10^{14}$ G). 

More creative interpretations of the period evolution (e.g. De Luca et al. 2004, Zavlin et al. 2004)
require scenarios such as bynary system or peculiar glitching.
However, such types of behaviour  are already
tightly constrained by the lack of an optical counterpart.

Are the unique characteristics of \ee pointing towards a new class of NSs?



\bibliographystyle{aa}

\end{document}